\documentclass[aps,prd,twocolumn,showpacs,eqsecnum,nofootinbib]{revtex4}

\usepackage{epsfig}
\usepackage{graphicx}
\usepackage{dcolumn}
\usepackage{amsmath}
\usepackage{enumerate}
\usepackage{epstopdf} 

\begin{document}

\title{%
\vskip-6pt \hfill {\rm\normalsize UCLA/04/TEP/19} \\
\vskip-12pt~\\
DAMA Dark Matter Detection Compatible with Other Searches}

\author{
\mbox{Graciela Gelmini$^{1}$} and
\mbox{Paolo Gondolo$^{2}$}}

\affiliation{
\mbox{$^1$  Department of Physics and Astronomy, UCLA,
 405 Hilgard Ave. Los Angeles, CA 90095, USA}
\mbox{$^2$ Department of Physics, University of Utah,
   115 S 1400 E \# 201, Salt Lake City, UT 84112, USA }
\\
{\tt gelmini@physics.ucla.edu},
{\tt paolo@physics.utah.edu}}

\date{May 27, 2004}

\vspace{6mm}
\renewcommand{\thefootnote}{\arabic{footnote}}
\setcounter{footnote}{0}
\setcounter{section}{1}
\setcounter{equation}{0}
\renewcommand{\theequation}{\arabic{equation}}

\begin{abstract} \noindent
We present two examples of velocity distributions for light dark matter particles
that reconcile the annual modulation signal
 observed by DAMA with all other negative results from dark matter
 searches. They are: (1) a conventional Maxwellian distribution for
 particle masses  of  6 to 9~GeV; (2) a dark matter stream coming from the
 general direction of Galactic rotation (not the Sagittarius stream).
 Our idea is based on having a signal in Na, instead of I, in DAMA,
 and can be tested in the immediate future by CDMS-II (using Si) and 
CRESST-II (using O).
\end{abstract}

\pacs{14.60.St, 98.80.Cq}

\maketitle

The nature of dark matter is one of the fundamental problems of physics and cosmology. 
Popular candidates for  dark matter are weakly interacting massive particles (WIMPs).
Direct searches for dark matter WIMPs  aim at detecting the scattering of
WIMPs off of nuclei in a low-background detector. These experiments measure
 the energy of the recoiling nucleus, and are sensitive to a signal above a 
detector-dependent energy threshold. 

One such experiment, the DAMA collaboration \cite{DAMA},  has found an annual
modulation in its data compatible with the signal expected from dark 
matter particles bound to our galactic halo \cite{FreeseDrukier}. Other such
experiments, such as CDMS \cite{CDMS-I,CDMS-II}, EDELWEISS
\cite{EDELWEISS}, and CRESST \cite{CRESST-I,CRESST-II}, have not found any
signal from WIMPs. It has been difficult to reconcile a WIMP signal in DAMA
 with the other negative results.

Here we address this issue by posing the following question: is there a  possible dark matter
signal above threshold for DAMA and under threshold for CDMS and EDELWEISS,
so that the positive and negative detection results would be compatible?
The answer is yes.

We assume that dark matter consists of WIMPs whose
 mass and cross section we choose in a purely phenomenological way, 
without any attempt of providing an elementary particle model to support them. 

The minimum dark matter particle velocity required to produce a certain nuclear
recoil energy $E$ is
\begin{equation}
v=\sqrt{\frac{ME}{2\mu^2}} = \sqrt{\frac{(m+M)^2 E}{2 M m^2}~,}
\label{v1}
\end{equation}
where $\mu = m M/(m+M)$ is the reduced WIMP-nucleus mass, $m$ is the WIMP
mass and $M$ is the nucleus mass. The nuclear energy 
threshold $E_{\rm th}$ observable with a particular nucleus corresponds to a minimum 
observable WIMP velocity, the velocity threshold $v_{\rm th}$. Threshold energies for
 several direct detection experiments are collected in Table 1.

To make our argument as simple as possible, consider for a moment the case 
 $m \ll M$. Then $\mu \simeq m$ is independent of the nucleus mass $M$, and 
$v_{\rm th}$ is proportional to $\sqrt{ ME_{\rm th}}$.
Using the nuclear masses of Na and Ge, $M_{\rm Na}= 21.41$~GeV  and 
  $M_{\rm Ge}= 67.64$~GeV, and the
energy thresholds in Table 1, 
the product $M E_{\rm th}$ is smaller for Na in DAMA than for Ge in CDMS-I.
For $m\ll M$, the $v_{\rm th}$ of CDMS-I is 2.44 times that of DAMA, and the
$v_{\rm th}$ of CDMS-II and EDELWEISS (as well as those of
other experiments using heavier nuclei) 
are  larger. Using the full Eq.~(\ref{v1}), it is easy to see that
 the velocity threshold of 
Na  in DAMA is smaller than that of Ge in CDMS-I for $m < 22.3~{\rm GeV}$.

This means that for light enough WIMPs it could be possible to have 
a virialized or non-virialized (a stream) halo component
  which could have a velocity above
threshold for Na in DAMA,  and below threshold for Ge in  CDMS
and EDELWEISS. That is, we could have a dark matter signal
visible for DAMA but not observable by CDMS and EDELWEISS.

CDMS has a small component of Si too.
Si is lighter than Ge, although  heavier than Na,
$M_{\rm Si}= 26.16$~GeV. Given the mentioned  nuclear energy recoil thresholds,
the velocity threshold of Si in CDMS-I is smaller than that of Na in DAMA for
all WIMP mass values.
However,  considering the CDMS efficiency close to 5~keV energies is 
about 10\%, the exposure of the Si detector of CDMS near threshold is 
about 0.3~kg-day, which may be too small
to have detected the signal which DAMA  might have seen though its Na.
In any event, CDMS has not yet used its Si component to set limits on dark matter, but
only to help in background rejection.

Light  nuclei are used by CRESST, in particular O, $M_{\rm O}= 14.90$~GeV.
CRESST-I~\cite{CRESST-I} used  sapphire (Al$_2$O$_3$), which besides O contains
 Al, similar in mass to Si. CRESST-I has set limits on dark matter with a very low nuclear 
recoil threshold of 0.6~keV, but with a small exposure of only
1.5~kg-day. The  velocity threshold for O in CRESST-I is so low that
CRESST-I is  sensitive to the bulk of the halo dark matter particles we are proposing.
CRESST-II uses calcium tungstate (CaWO$_4$), which also contains the light
O nucleus, but background discrimination sets a relatively high threshold
of $\sim 10$ keV. CRESST-II has run a prototype without neutron shield and
set  the limits quoted in Table 1 \cite{CRESST-II}. 
The completed CRESST-II will test our idea.

\begin{figure}
\includegraphics[width=0.45\textwidth]{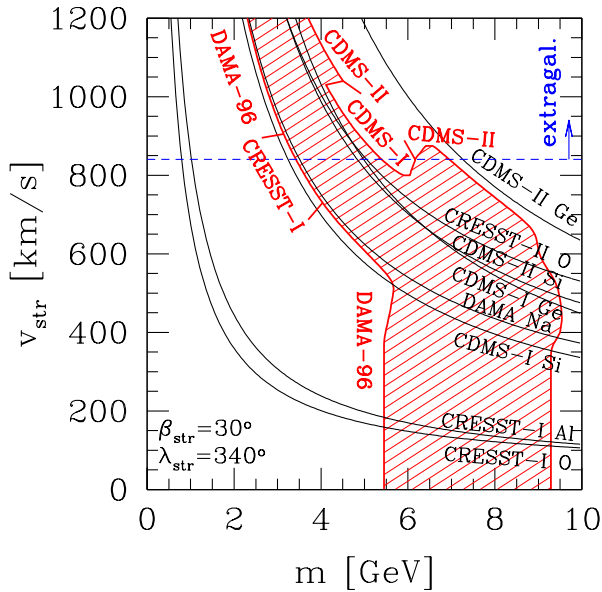}
\vspace{-5pt}
\caption{Stream heliocentric speed $v_{\rm str}$  (here assumed to arrive from
ecliptic latitude $\beta_{\rm str}=30^\circ$ and ecliptic longitude
 $\lambda_{\rm str}=340^\circ$) vs.\
WIMP mass $m$. The hatched region is compatible with DAMA,
 CDMS, EDELWEISS, and CRESST. Which experiment limits the compatible
 region is indicated along its edges. Also marked are the speed above which the stream
is extragalactic (dashed horizontal line), and the threshold speeds 
$v_{\rm th}$ of several experiments and target nuclei.}
\vspace{-5pt}
\end{figure}

Fig.~1 shows the relevant velocity thresholds $v_{\rm th}$, as functions of the
WIMP mass $m$. Our idea is that
a contribution to the density of WIMPs with
 velocities smaller than the CDMS threshold
but larger that the DAMA threshold could 
explain the data. We are here thinking
in terms of a  non-virialized stream of dark matter~\cite{stream}
 added to an isothermal halo model. The stream would
be the dominant signal in DAMA (because there would be few halo WIMPs
above the DAMA threshold) while  it would change the signal due to the 
bulk of the halo (e.g.\ in CRESST) very
little. The vertical axis of  Fig.~1 gives the average 
dark matter stream  velocity relative to the Sun. Assuming a 
stream as described below,
the region where our idea works is the hatched region in the 
figure. To our surprise, the allowed region reaches zero stream velocity
in a small interval of WIMP masses near 6 to 9~GeV. In this interval, a
stream is not necessary and just the contribution of an isothermal halo
would do. 

If there is no halo contribution above threshold in DAMA, the stream
 arrival direction is limited by the requirement that the DAMA 
modulation peaks  May 21 $\pm$ 22 days. May 21 is 61 days after the
 Spring equinox (March 21), and thus the Sun is at ecliptic longitude
 $61/365.25 \times 360=60^\circ$. Since the radius vector to the Sun
 and the velocity of the Earth are perpendicular (for a circular orbit),
 the Earth  is moving toward a point of ecliptic longitude 
$60^\circ-90^\circ=330^\circ$. Thus, in case the DAMA modulation 
is entirely due to a dark matter stream, the arrival direction of 
the stream on Earth must have ecliptic longitude 
$\lambda=330^\circ \pm 22^\circ$. 
The amplitude of the modulation depends on the projection 
of the stream velocity onto the ecliptic, and is proportional
 to $\cos\beta_{\rm str}$, where $\beta_{\rm str}$ is the ecliptic 
latitude of the stream arrival direction.

The non-virialized  dark matter stream  we are proposing 
 could be bound  to our galaxy or not. 
There are non-virialized dark matter streams bound to our own galaxy, such
as the tidal streams of the Sagittarius dwarf galaxy \cite{Sgr}. The
Sagittarius leading tidal stream passes through the solar neighborhood,
with a heliocentric speed of $\approx 350$~km/s, 
but its arrival direction $(\lambda,\beta)\approx (187^\circ,8^\circ)$
does not have the correct ecliptic longitude to give the observed phase 
of the DAMA modulation without a substantial contribution from the usual 
halo component. So the stream
we are implying should be a different, yet undiscovered, stream.
 There may be dark matter bound not to our galaxy 
but to our Local Group of galaxies~\cite{localg}, and also 
 dark matter bound to our supercluster, possibly passing through us~\cite{FGS}.
Its galactocentric incoming velocity $v_{\rm in}$
is increased by gravitational focusing while falling into our
galaxy to become a galactocentric velocity 
$v_{\rm local} =\sqrt{v_{\rm in}^2 +v_{\rm esc}^2}$ near the Sun. Here $v_{\rm esc}$ is the local escape velocity from the Galaxy.
 The density of an incoming
 stream is also increased by focusing, at least linearly with the ratio
$v_{\rm local}/{v_{\rm in}}$, but possibly by much larger factors, which are
however complicated to evaluate. Interestingly, the Sun moves relative 
to the Local Group at $306\pm 18$~km/s in direction 
$(\lambda,\beta)=(4^\circ\pm10^\circ, 57^\circ \pm
4^\circ)$~\cite{Courteau},  the ecliptic longitude of which is not far from the 
desired value.
\begin{table*}
\begin{tabular}{||l|l|l|l|l|c||}
\hline\hline
Experiment & Exposure [kg-day] & Threshold [keV] & Efficiency [\%] & Constraint & Ref. \\
\hline
CDMS-I & \begin{minipage}{50pt} \begin{flushleft} Si: 2.83\\Ge: 28.3\end{flushleft} \end{minipage} & 5 &  \begin{minipage}{70pt} \begin{flushleft} $E<10$keV: 7.6 \\ $E<20$keV: 22.8 \\ $E>20$keV: 38 \end{flushleft} \end{minipage} & 5--55keV: $<$2.3 events ($\dagger$) & \protect\cite{CDMS-I} \\
\hline
CDMS-II & \begin{minipage}{50pt} \begin{flushleft}Si: 5.26\\Ge: 52.6 \end{flushleft} \end{minipage} & 10 &  \begin{minipage}{70pt} \begin{flushleft}  $E<20$keV: 22.8 \\ $E>20$keV: 38 \end{flushleft} \end{minipage} & 10--100keV: $<$2.3 events ($\dagger$) & \protect\cite{CDMS-II} \\
\hline
EDELWEISS & Ge: 8.2 & 20 &  100 & 20--100keV: $<$2.3 events  ($\dagger$) & \protect\cite{EDELWEISS} \\
\hline
CRESST-I & Al$_2$O$_3$: 1.51 & 0.6 &  100 &  ($\ddag$) & \protect\cite{CRESST-I} \\
\hline
CRESST-II & CaWO$_4$: 10.448 & 10 &  100 & \begin{minipage}{180pt} \begin{flushleft} Ca+O, 15--40keV: $<$6 events  \\ W, 12--40keV: $<$2.3 events ($\dagger$) \end{flushleft} \end{minipage}  & \protect\cite{CRESST-II} \\
\hline
DAMA/NaI-96 & NaI: 4123.2 & \begin{minipage}{50pt} \begin{flushleft} I: 22 ($\diamond$) \\ Na: 6.7 ($\diamond$)  \end{flushleft} \end{minipage}  &  100 &  \begin{minipage}{180pt} \begin{flushleft} 1--2keVee: $<$1.4/kg-day-keVee ($\star$) \\ 2--3keVee: $<$0.4/kg-day-keVee ($\star$)  \end{flushleft} \end{minipage}  & \protect\cite{DAMA96} \\
\hline
DAMA/NaI-03 & NaI: 107731 & \begin{minipage}{50pt} \begin{flushleft} I: 22 ($\diamond$) \\ Na: 6.7 ($\diamond$)  \end{flushleft} \end{minipage} &  100 & 2--4keVee: 0.0233$\pm$0.0047/kg-day-keVee ($\bullet$) & \protect\cite{DAMA} \\
\hline\hline
\end{tabular} 
\caption{Experimental constraints used in this study. Notes to the table: ($\dagger$) upper limit assuming no detected event; 
($\ddag$) to reproduce the published curve in~\protect\cite{CRESST-I}, we impose appropriate upper limits all along the recoil spectrum in their Fig.~1;
($\diamond$) from an electron
equivalent threshold of 2~keVee,  using the quenching factors $Q=E_{\rm ee}/E$ equal to 0.09 for I and  0.3 for Na~\protect\cite{DAMA};
($\star$) approximations that reproduce the published $\sigma_{\rm p}$ vs.\ $m$ limit across our mass range; 
($\bullet$) amplitude of annual modulation.}
\vspace{-5pt}
\end{table*}

Our procedure is the following.
For the WIMP velocity distribution, we take  $f({\bf v},t) =
 f_{\rm h}({\bf v},t) + f_{\rm str}({\bf v},t),$ the sum of a halo 
distribution $f_{\rm h}({\bf v},t)$ and an optional contribution from
 a dark matter stream $f_{\rm str}({\bf v},t)$. For $f_{\rm h}({\bf v},t)$
 we take the conventional truncated Maxwellian used in the comparison of 
direct detection experiments,
\begin{equation}
f_{\rm h}({\bf v},t) =
 \frac{1}{N_{\rm h}(\pi {\overline v}_{\rm h}^2)^{3/2}} e^{- | {\bf v}+
{\bf v}_{\odot}+{\bf v}_{\oplus}(t) |^2/{\overline v}_{\rm h}^2 } ,
 \end{equation}
for $ | {\bf v}+{\bf v}_{\odot}+{\bf v}_{\oplus}(t) | < v_{\rm esc},$
 and $ f_{\rm h}({\bf v},t) =0 $ otherwise.
Here $N_{\rm h} = {\rm erf}(z)-2\pi^{-1/2}ze^{-z^2}$, with
 $z=v_{\rm esc}/{\overline v}_{\rm h}$, is a normalization factor,
 ${\bf v}_{\odot}$ is the velocity of the Sun relative to the Galaxy 
(galactocentric velocity), and
 ${\bf v}_{\oplus}(t)$ is the velocity of the Earth relative to the Sun. 
For ${\bf v}_{\odot}$, we take the conventional value $v_{\odot} = 232$
 km/s, which is within the measured value of $233\pm3$~km/s in the 
direction  $(\lambda,\beta) = (341^\circ \pm 1^\circ,
 60.5^\circ \pm 0.5^\circ)$ \cite{solarmotion}. 
For ${\bf v}_{\oplus}$, we assume a magnitude of 29.8 km/s along a circular orbit on a
 plane inclined by $60^\circ$ with respect to ${\bf v}_{\odot}$.
Furthermore, we take the conventional velocity dispersion 
$ {\overline v}_{\rm h} = 220 $ km/s and the escape speed from 
the galaxy $v_{\rm esc} = 650$ km/s.
We make no claim that this is a realistic velocity distribution, 
but it offers us a definite benchmark for comparison. In addition, 
we assume the conventional value for the local dark matter density $\rho=0.3$ GeV/cm$^3$.
With the halo model we assume, the maximum possible heliocentric velocity of a halo
particle is  $v_{\rm esc} + v_{\odot}= 882$~km/s.

 For the stream,  we assume
\begin{equation}
f_{\rm str}({\bf v},t) =  \frac{\xi_{\rm str}}{(\pi 
{\overline v}_{\rm str}^2)^{3/2}} e^{- | {\bf v}- {\bf v}_{\rm str} +
 {\bf v}_{\oplus}(t) |^2/{\overline v}_{\rm str}^2 } ,
\end{equation}
where $\xi_{\rm str}$ is the local dark matter density in the stream
 (in units of 0.3 GeV/cm$^3$), ${\bf v}_{\rm str}$ is the stream
 velocity relative to the Sun (heliocentric velocity), and 
${\overline v}_{\rm str}$ is the velocity dispersion in the stream
 (which we fix at 20 km/s; our results depend very little on the value
of ${\overline v}_{\rm str}$).  To produce Figs.~1 and 2, we take
 $\xi_{\rm str}=0.03$ and a stream arrival direction  
 $(\lambda_{\rm str},\beta_{\rm str})=(340^\circ,30^\circ)$, 
so that the stream is at 30$^\circ$ of the
 ecliptic and at  30$^\circ$ of the Sun's galactocentric velocity.
Such a stream is extragalactic if its heliocentric velocity 
  exceeds  840~km/s (marked by the dashed horizontal line in Fig.~1).
Due to gravitational focusing, to obtain $v_{\rm local}> 660$~km/s, for example, i.e.\ a stream heliocentric radial
 velocity $v_{\rm str}> 892$~km/s, requires 
$v_{\rm in} > 114$~km/s,  a reasonable value.

For each choice of WIMP velocity distribution, we must make sure that
 we not only produce the correct amplitude for the DAMA modulation in
 the viable region but also that all of the current experimental
 constraints are satisfied. 
We consider constraints from DAMA/NaI-96~\cite{DAMA96}, 
DAMA/NaI-03~\cite{DAMA},  CDMS-I~\cite{CDMS-I},
 CDMS-II \cite{CDMS-II}, CRESST-I~\cite{CRESST-I}, and CRESST-II~\cite{CRESST-II}.
 The exposures, efficiencies, thresholds, and constraints
  we use are listed in Table 1.

We compute the expected number of recoil events with recoil energy
 in the range $(E_1,E_2)$ as 
\begin{equation}
N = \sum_i \int_{E_1}^{E_2} \frac{dR_i}{dE} \, {\cal E}_i(E) \, d E.
\end{equation}
The sum runs over the nuclear species in the detector,
 and ${\cal E}_i = {\cal M}_i T_i \epsilon_i(E)$
 is the effective exposure of species $i$ ($T_i$ being 
the time the mass ${\cal M}_i$ is exposed to the signal, and $\epsilon_i(E)$ being
 the efficiency). Moreover, $dR_i/dE$
 is the expected recoil rate per unit detector mass and unit time,
\begin{equation}
\frac{dR_i}{dE} = 
\frac{ \rho \sigma_{i} | F_i(E) |^2 }{ 2 m \mu_{i}^2 } \int_{v \!>\!
 \sqrt{M_i E/2 \mu_{i}^2 }} \frac{f({\bf v},t)}{v} \, d^3v .
\end{equation}
Here $M_i$ is the nuclear mass, $\mu_i = m M_i/(m+M_i)$ is the reduced
 WIMP-nucleus mass, $\rho$ is the local halo WIMP density, $F_i(E)$ is
 a nuclear form factor, $\sigma_i$ is the WIMP-nucleus cross section,
 and $f({\bf v},t)$ is the WIMP velocity distribution in the reference
 frame of the detector.

In this analysis, we assume that the WIMP-nucleus interaction is 
spin-independent and scales with the square of the nucleus
 atomic number $A_i$ as $\sigma_i = \sigma_{\rm p} A_i^2 
(\mu_i/\mu_{\rm p})^2$. 
 Here $\sigma_{\rm p}$ is the WIMP-proton cross section.
 For the nuclear form factor we use the conventional Helmi form,
 $F_i(E) = 3 e^{-q^2s^2/2} [\sin(qr)- qr\cos(qr)]/(qr)^3,$ with $s=1$~fm,
 $r=\sqrt{R^2-5s^2}$, $R=1.2 A^{1/3}$~fm, $q=\sqrt{2 M_i E}$.

To find a compatible region, we vary the WIMP mass, the stream density, and
the stream velocity (allowing also for the absence of the stream). For
each choice of these variables we find the WIMP-proton cross section 
$\sigma_{\rm p}$ that produces the desired modulation amplitude in
the 2-4 keVee bin of DAMA/NaI-03 within two sigma (the resulting
$\sigma_{\rm p}$ are shown in Fig.~2; we tried other ways to fix
$\sigma_{\rm p}$ using also the 2-5, and 2-6 keVee bins and obtained
similar results). With this WIMP-proton cross section, we evaluate the 
expected number of events in all of the other experiments. We thus
determine if the parameters we choose are compatible with the experimental 
constraints we impose. 

Fig.~1 shows the compatible region in heliocentric stream speed versus 
WIMP mass resulting from our study, cut arbitrarily to  $v_{\rm str} < 1200$~km/s.
The region is bounded from above by the constraints from CDMS-I and II, 
 and from below by the constraint from DAMA/NaI-96 and CRESST-I, up to
 masses of about 6~GeV.
The vertical band around  $m=6$ to 9~GeV shows that
there the stream speed can be arbitrarily small. Indeed, in that mass range,
the DAMA modulation can be reproduced with just 
the conventional Maxwellian distribution, without an additional stream.

The hatched region in Fig.~2 shows the values of the   WIMP-proton cross
section $\sigma_{\rm p}$ resulting from our study that are compatible with the
DAMA/NaI-03 modulation,
as a function of the WIMP mass. The region was cut arbitrarily at 
  $v_{\rm str} < 1200$~km/s, as in Fig~1. Also shown is where the cut would be for a Galactic stream, $v_{\rm str} < 840$~km/s.
The cross-hatched  region is the case of a conventional Maxwellian halo without a stream.
The upper bounds on  $\sigma_{\rm p}$ are given by CRESST-I, CDMS-I and II, and DAMA/NaI-96.
The values we find are included in the region presented as allowed by DAMA
 in Fig.~28 of Ref.~\cite{DAMA}.  Light neutralinos as WIMPs with masses as
 low as 2~GeV \cite{stodolsky} or, with updated bounds, 6~GeV \cite{bottino} 
 have been considered, but their cross
 sections are about one order of magnitude smaller than those needed here.

\begin{figure}
\includegraphics[width=0.45\textwidth]{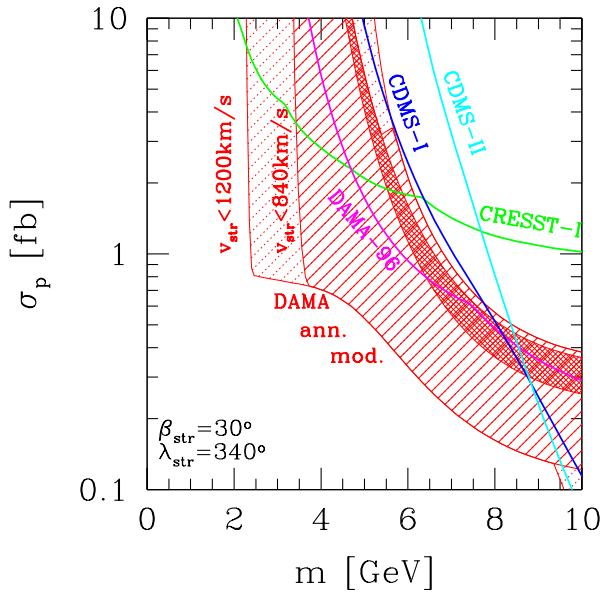}
\vspace{-5pt}
\caption{Region of WIMP-proton cross section $\sigma_{\rm p}$ resulting
  from our study  vs.\ WIMP mass $m$ 
(cross-hatched region: without stream; hatched region: with stream and velocity cuts as shown). Also limits from
DAMA/NaI-96, CRESST-I and II, and CDMS-I and II (see Table
1). }
\vspace{-5pt}
\end{figure}

In conclusion, we have pointed out that for light dark matter particles
a signal could be observed by DAMA through its Na component. This signal
would be below threshold for Ge in CDMS and EDELWEISS.
This possibility can be
tested with a few months of Si data in CDMS-II, and
future O data in CRESST.

For WIMPs with spin independent interactions, 
we have presented two examples of dark matter velocity distributions
that give the annual modulation observed by DAMA but satisfy all other
constraints from dark matter searches. The first is a 
conventional Maxwellian distribution with a WIMP mass around 6 to 9~GeV.
Surprisingly, this simple possibility remains open.
Our second example is the former distribution superposed to
a dark matter stream coming from the general 
direction of the Galactic rotation (perhaps
 associated with extragalactic dark matter, but {\it not} the 
Sagittarius stream). 

For the sake of illustration, we have assumed  a particular
 density of the stream (0.03 of the local halo density) and a particular incoming direction 
 (halfway between the plane of the Earth's orbit and the
 direction of the Sun's velocity in the Galaxy).
Our main result are the allowed regions presented in Figs.~1 and 2. 

Clearly, the effect of the stream would be larger (smaller) for
 larger (smaller) stream  densities and for
incoming directions closer to (further from) the plane of the Earth's orbit.  
For simplicity, we have illustrated our idea only for 
the case of WIMPs with spin-independent interactions.
Other kinds of particles and interactions, or  
halo velocities distributions more complicated than a conventional Maxwellian distribution,
 may extend the allowed regions of parameters.

This work was supported in part by the US Department of Energy Grant
DE-FG03-91ER40662, Task C and NASA Grant NAG5-13399.

\end{document}